\documentclass[12pt,preprint]{aastex}

\newcommand{\GOES}{{\em GOES}}
\newcommand{\SDO}{{\em SDO}}
\newcommand{\Hinode}{{\em Hinode}}
\newcommand{\feiv}{Fe {\sc xxiv}}
\newcommand{\feiii}{Fe {\sc xxiii}}
\newcommand{\fevi}{Fe {\sc xvi}}

\shorttitle{Spectroscopic Observations of Current Sheet}
\shortauthors{Li et al.}

\begin{document}

\title{Spectroscopic Observations of a Current Sheet in a Solar Flare}

\author{\small{Y. Li$^{1,2,3}$, J. C. Xue$^{1}$, M. D. Ding$^{2}$, X. Cheng$^{2}$, Y. Su$^{1}$, L. Feng$^{1}$, J. Hong$^{2}$, H. Li$^{1}$, W. Q. Gan$^{1}$}}
\affil{$^1$Key Laboratory of Dark Matter and Space Astronomy, Purple Mountain Observatory, Chinese Academy of Sciences, Nanjing 210034, China; yingli@pmo.ac.cn}
\affil{$^2$School of Astronomy and Space Science, Nanjing University, Nanjing 210023, China}
\affil{$^3$CAS Key Laboratory of Solar Activity, National Astronomical Observatories, Beijing 100012, China}

\begin{abstract}
Current sheet is believed to be the region of energy dissipation via magnetic reconnection in solar flares. However, its properties, for example, the dynamic process, have not been fully understood. Here we report a current sheet in a solar flare (SOL2017-09-10T16:06) that was clearly observed by the Atmospheric Imaging Assembly on board the {\em Solar Dynamics Observatory} as well as the EUV Imaging Spectrometer on \Hinode. The high-resolution imaging and spectroscopic observations show that the current sheet is mainly visible in high temperature ($>$10 MK) passbands, particularly in the \feiv~192.03 \AA~line with a formation temperature of $\sim$18 MK. The hot \feiv~192.03 \AA~line exhibits very large nonthermal velocities up to 200 km s$^{-1}$ in the current sheet, suggesting that turbulent motions exist there. The largest turbulent velocity occurs at the edge of the current sheet, with some offset with the strongest line intensity. At the central part of the current sheet, the turbulent velocity is negatively correlated with the line intensity. From the line emission and turbulent features we obtain a thickness in the range of 7--11 Mm for the current sheet. These results suggest that the current sheet has internal fine and dynamic structures that may help the magnetic reconnection within it proceeds efficiently.

\end{abstract}


\noindent

\noindent

\section{Introduction}

Solar flares are one of the most energetic events in the solar atmosphere, which are often accompanied by coronal mass ejections (CMEs) (see a recent review by \citealt{schm15}). In the standard flare/CME model, a magnetic flux rope (MFR) suffering loss of equilibrium rises and stretches a thin and long-extended current sheet that connects the overlying MFR (or CME) and the flare loops underneath \citep{linj00,link03}. The current sheet is believed to be the energy dissipation region, where magnetic reconnection takes place powering the flare and accelerating the CME (e.g., \citealt{prie02}). 

Directly observing the current sheet in solar eruptions is very difficult because of its tenuous emission and narrow width or thickness \citep{linj15}. Nevertheless, some imaging and also spectroscopic observations have revealed the evidence of current sheet. \cite{webb03} found a set of ray-like structures form under the CME and regarded as signatures of the current sheet in white light. \cite{koyk03} and \cite{linj05} interpreted the thin streamers above a cusp structure as the current sheet formed behind the CME. Recently, using the high-resolution EUV images from the Atmospheric Imaging Assembly (AIA; \citealt{leme12}) on board the {\em Solar Dynamics Observatory} (\SDO), some current-sheet-associated structures were observed as rays when viewed edge on or as fans when viewed face on \citep{reev11,warr11,sava12,mcke13}. In addition, using the Ultraviolet Coronagraph Spectrometer (UVCS) on the {\em Solar and Heliospheric Observatory} ({\em SOHO}), \cite{ciar02} detected a narrow bright feature between CME core and flare arcades at a heliocentric distance of 1.5 R$_\sun$ in the hot Fe {\sc xviii} 974 \AA~line and suggested that the bright feature could be the region where the reconnecting current sheet was located. Some samples of bright rays detected in the UVCS spectra also support these narrow bright features as current sheets \citep{ciar13}.

The physical and dynamical properties of current sheets were investigated in the past twenty years. For example, the temperature inside the current sheet was measured to be a few MK based on the line ratio technique \citep{ciar02,koyk03,bemp06,ciar08}. From the line broadening, nonthermal velocities in the current sheet can be measured. \cite{ciar02} obtained nonthermal velocities of 60 km s$^{-1}$ in the Fe {\sc xviii} line that were interpreted as turbulent motions in the current sheet. \cite{bemp08} also derived turbulent velocities of the order of $\sim$60 km s$^{-1}$ in the Fe {\sc xviii} line, which slowly decayed down to $\sim$30 km s$^{-1}$ subsequently. Moreover, nonthermal velocities of 30--200 km s$^{-1}$ were detected in the current sheet, also indicative of the presence of turbulence \citep{ciar08,susi13,dosc14}.

Thickness of the current sheet is another important parameter, which may reflect the property of dynamics in the current sheet. Adopting the full width at half-maximum (FWHM) of the Fe {\sc xviii} emission profile along the UVCS slit as the thickness of current sheet, \cite{ciar08} found that the apparent thickness changed from 140 to 70 Mm during the eruption, which was similar to the ones in some other events also observed by UVCS \citep{ciar02,koyk03}. Using the Large Angle and Spectrometric Coronagraph (LASCO) on board {\em SOHO}, \cite{linj09} inferred the thickness of current sheet increasing from 50 to 450 Mm in one event and from 100 to 750 Mm in another event. Moreover, \cite{sava10} reported an average thickness of 4--5 Mm for the candidate current sheet using the images from \Hinode's X-ray Telescope, which was similar to the results obtained from AIA images \citep{seat17}. Note that all these thickness values of current sheets are much greater than the theoretical value of only tens of meters \citep{litv96,wood05}. 

In this Letter, we present excellent spectroscopic observations of a current sheet from the EUV Imaging Spectrometer (EIS; \citealt{culh07}) on \Hinode. The current sheet is clearly visible in the \feiv~192.03 \AA~line with a formation temperature of $\sim$18 MK. The hot \feiv~line exhibits very large nonthermal velocities up to 200 km s$^{-1}$ in the current sheet, indicative of the presence of turbulence. The largest turbulent velocity is located at the edge of the current sheet and is not cospatial with the greatest line intensity. From the line emission and turbulent features, we obtain a thickness of 7--11 Mm for the current sheet. These results can help understand the nature of the current sheet and the physical processes that occur inside the current sheet.

\section{Observations and Data Reduction}
\label{sec-obs}

\begin{figure*}
\centering
\includegraphics[width=13cm]{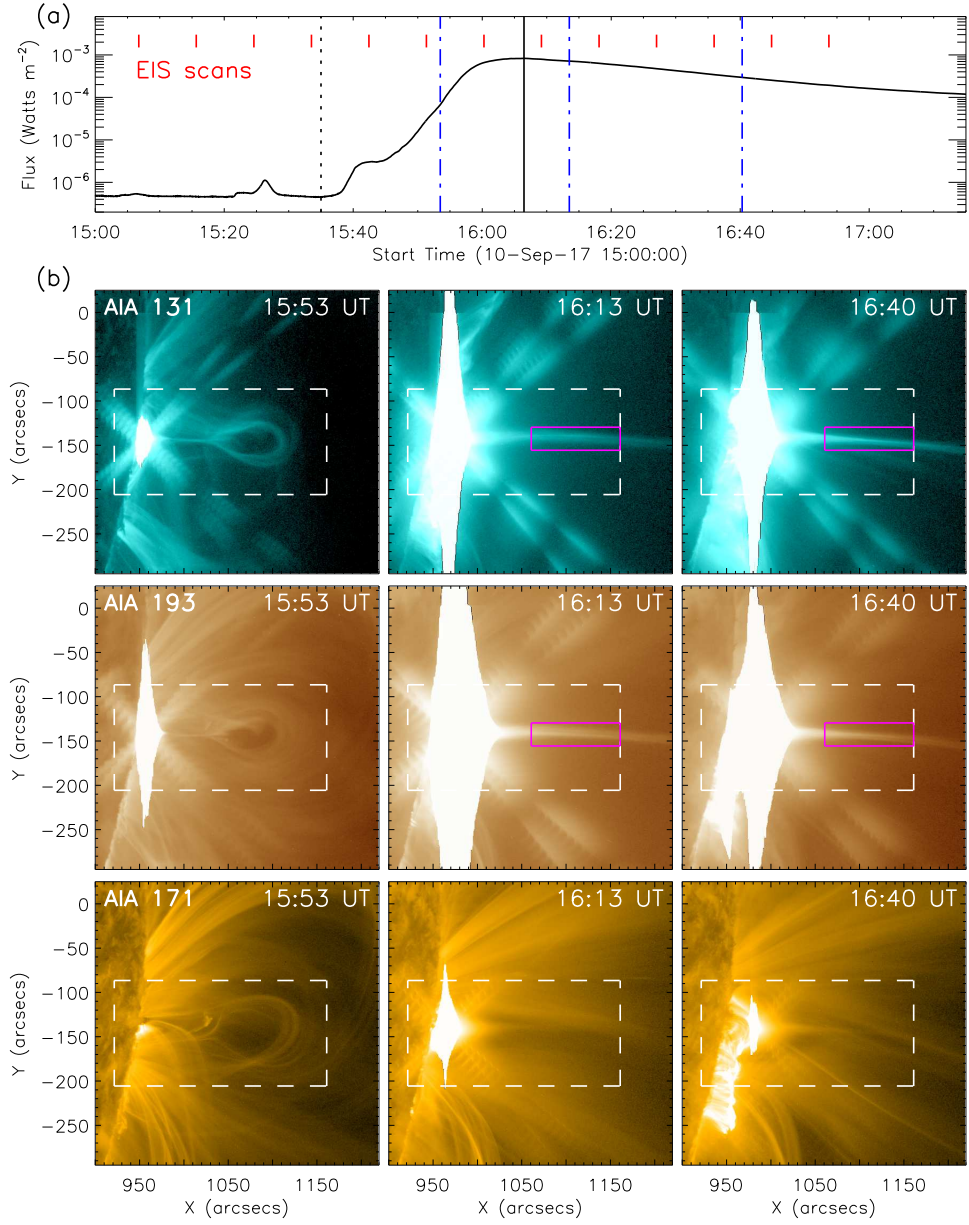}
\caption{{\small Overview of the flare observations. (a) \GOES~1--8 \AA~soft X-ray flux. The short vertical red lines indicate the EIS scans. The vertical black dotted and solid lines denote the flare onset and peak times, respectively. The three blue dashed-dotted lines mark the times of AIA images in panel (b). (b) AIA 131, 193, and 171 \AA~images during the flare. The white box denotes the region as shown in Figures \ref{fig-eis} and \ref{fig-vel}. The magenta box indicates the current sheet.}}
\label{fig-aia}
\end{figure*}

The event presented here is a \GOES~X8.2 flare that occurred at the west limb on 2017 September 10. The flare started at 15:35 UT and peaked at 16:06 UT (Figure \ref{fig-aia}(a)). In this flare, a bubble-like structure (probably an MFR) showed up at $\sim$15:50 UT (left panels in Figure \ref{fig-aia}(b)) and rose higher and higher over time. Behind it, a linear bright feature was stretched out, which is most likely a current sheet when viewed edge on, as stated in the standard flare/CME model. The long thin current sheet structure (see the magenta box in Figure \ref{fig-aia}(b)) was well observed by \SDO/AIA as well as \Hinode/EIS especially at its early evolutionary stage.

AIA observes EUV images with a high spatial resolution of $1.\!\!^{\prime\prime}2$ (or $0.\!\!^{\prime\prime}6$ pixel$^{-1}$) and a cadence of 12 s. The EUV channels include 193 \AA~(with emissions at $\sim$18 and $\sim$1.6 MK), 131 \AA~($\sim$11 and $\sim$0.4 MK), 94 \AA~($\sim$7.1 and $\sim$1.1 MK), 335 \AA~($\sim$2.5 MK), 211 \AA~($\sim$2.0 MK), and 171 \AA~($\sim$0.6 MK). In this event, the long thin current sheet is more clearly visible in the 193 and 131 \AA~passbands compared with the other channels (Figure \ref{fig-aia}(b)), implying that the current sheet has a high-temperature ($>$10 MK) emission. The differential emission measure (DEM) analysis of the same event also reveals that the temperature in the current sheet is $\sim$20 MK \citep{chen17,warr17}.

EIS obtains EUV spectra ($\sim$170--210 \AA~and $\sim$250--290 \AA) with a spectral resolution of 0.0223 \AA~pixel$^{-1}$. The data are corrected for dark current, hot pixels, and cosmic-ray hits via the standard EIS routine, yielding a physical unit in the line intensity. For this flare, EIS started a scan-mode observation before the flare onset, which lasted until 16:53 UT (see the short vertical lines in Figure \ref{fig-aia}(a)). The 2\arcsec~slit was used to scan an area of 240\arcsec$\times$304\arcsec~with a step of 3\arcsec~in the {\em X} direction and a scale of 1\arcsec~in the {\em Y} direction. The cadence for each run is $\sim$9 minutes.

\begin{figure*}
\centering
\includegraphics[width=16cm]{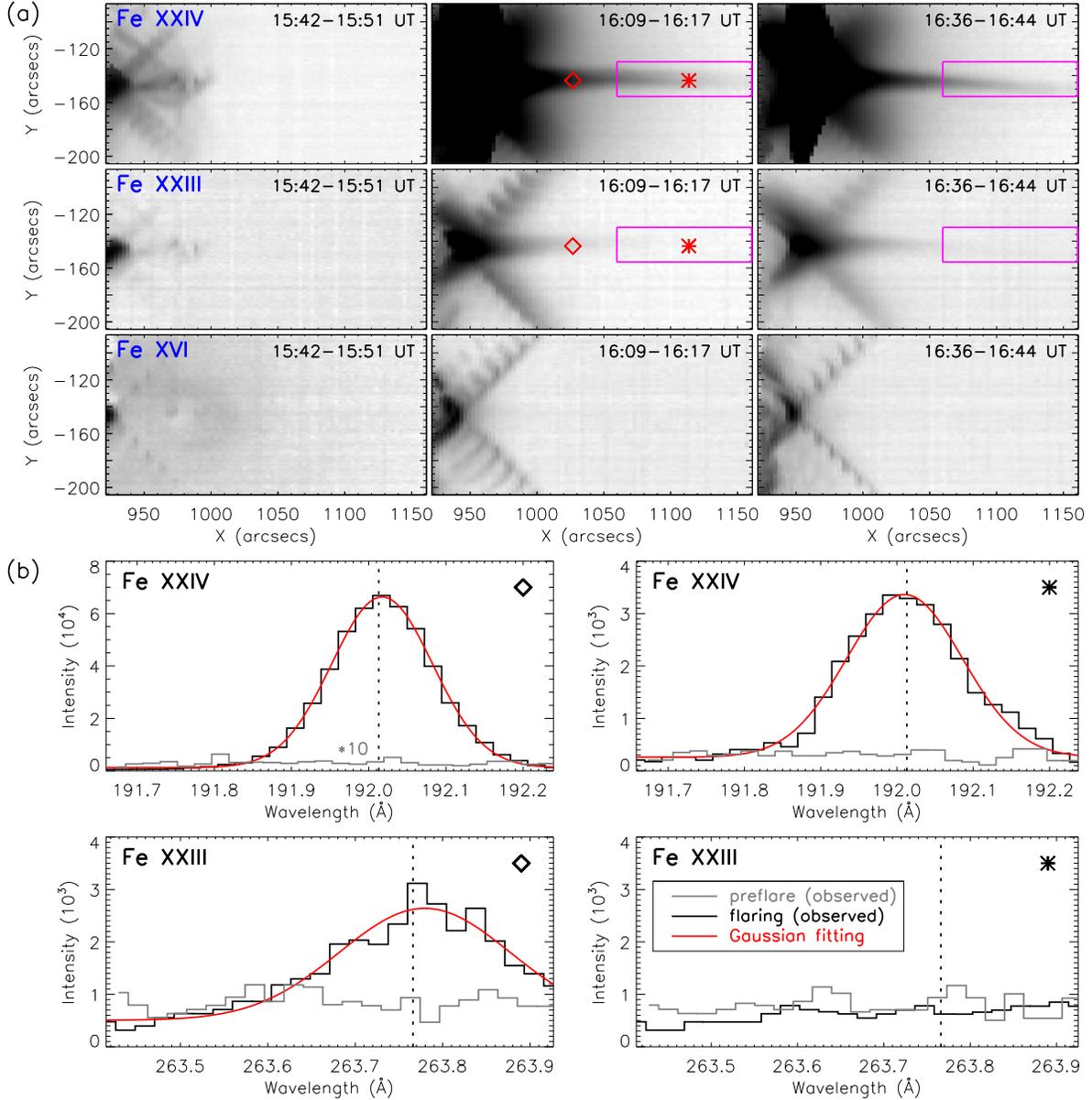}
\caption{{\small (a) Spectral images (in reversed colors) obtained from slit scans in the \feiv~($\sim$18 MK), \feiii~($\sim$14 MK), and \fevi~($\sim$3 MK) lines. The magenta box (same as in Figure \ref{fig-aia}) indicates the current sheet particularly in the \feiv~images. The diamond and asterisk symbols denote two sample locations (at the cusp and inside the current sheet, respectively) where the \feiv~and \feiii~line profiles are plotted in panel (b). (b) Observed \feiv~and \feiii~line profiles (black and gray curves) at the sample locations and the single Gaussian fittings (red curves). The vertical dotted line represents the line center. Note that the preflare \feiv~profile (gray) at the location marked by diamond is multiplied by 10. It is seen that no obvious \feiii~emission appears at the location marked by asterisk.}}
\label{fig-eis}
\end{figure*}

EIS observed the flare in tens of spectral lines with a wide temperature range of 0.2--18 MK. The current sheet structure is mainly visible in the hottest \feiv~192.03 \AA~line ($\sim$18 MK) (Figure \ref{fig-eis}(a)), which is consistent with AIA observations. So we focus on this line in the present study. Note that the \feiv~192.03 \AA~line is blended with the Fe {\sc xi} 192.02 \AA~line ($\sim$1.2 MK). However, the line blending can be safely ignored in this flare especially for the current sheet region by two reasons. (1) Theoretically, the Fe {\sc xi} 192.02 \AA~line is much weaker than the Fe {\sc xii} 192.39 \AA~line which lies in the same spectral window but is unblended with the \feiv~192.03 \AA~line. If the observed emission at 192.02/192.03 \AA~is stronger than the Fe {\sc xii} 192.39 \AA~line, it should be mainly contributed by the \feiv~192.03 \AA~line. This is just the case here. (2) We use the temperature and density derived from DEM analysis to calculate the intensities of \feiv~192.03 \AA~and Fe {\sc xi} 192.02 \AA~lines. The result shows that there is almost no Fe {\sc xi} 192.02 \AA~but very strong \feiv~192.03 \AA~emission in the current sheet region. In addition, a relatively cooler line of \feiii~263.76 \AA~($\sim$14 MK) is used to determine the lower end of the current sheet (i.e., the cusp tip). Here we assume that the current sheet itself could only appear in the \feiv~line while the cusp (or outflowing region) could be seen in both of the \feiv~and \feiii~lines (see the line profiles in Figure \ref{fig-eis}(b)). This seems to be reasonable since the current sheet itself has a relatively lower density and a more uniform thickness compared with the cusp. Thus, the lower end of the current sheet is estimated to be located at a height of $\sim$75 Mm, i.e., the left boundary of the magenta box as plotted in Figures \ref{fig-aia}--\ref{fig-vel}. Note that the height of the lower end might be underestimated, but it should not affect our results substantially.

We use a single Gaussian function to fit the \feiv~line profiles. From Figure \ref{fig-eis}(b), it is seen that the fitting curves match the observed profiles very well. When calculating the nonthermal velocity, we subtract the instrumental width of EIS (0.056 \AA) and the thermal width at 18 MK (73 km s$^{-1}$) from the \feiv~line broadening. We also use the temperature derived from DEM analysis and find a quite similar result for the nonthermal velocity.

\section{Results}

\subsection{Spatio-temporal Variations of Line Intensity and Nonthermal Velocity}

\begin{figure*}
\centering
\includegraphics[width=14cm]{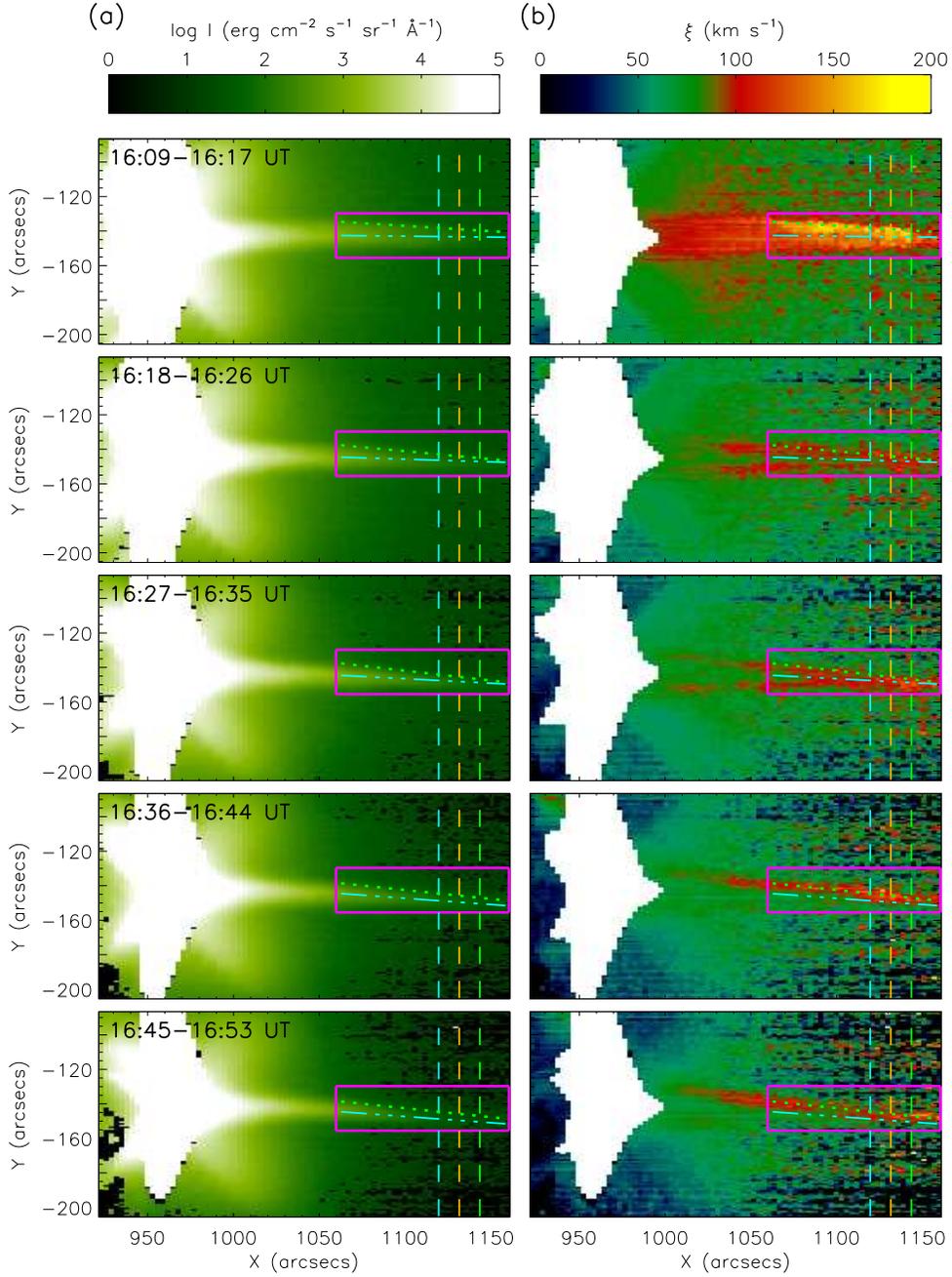}
\caption{{\small Temporal evolution of the \feiv~line intensity (a) and nonthermal velocity (b). The magenta box (same as in Figure \ref{fig-eis}) indicates the current sheet region. The green dotted and cyan dashed-dotted lines mark the locations where the \feiv~line intensity and nonthermal velocity are plotted in Figure \ref{fig-htm}. The three vertical dashed slices are used to measure the thickness of the current sheet, as shown in Figure \ref{fig-wid}. Note that the white area especially in panel (b) represents a saturation in the \feiv~line profile.}}
\label{fig-vel}
\end{figure*}

\begin{figure*}
\centering
\includegraphics[width=14cm]{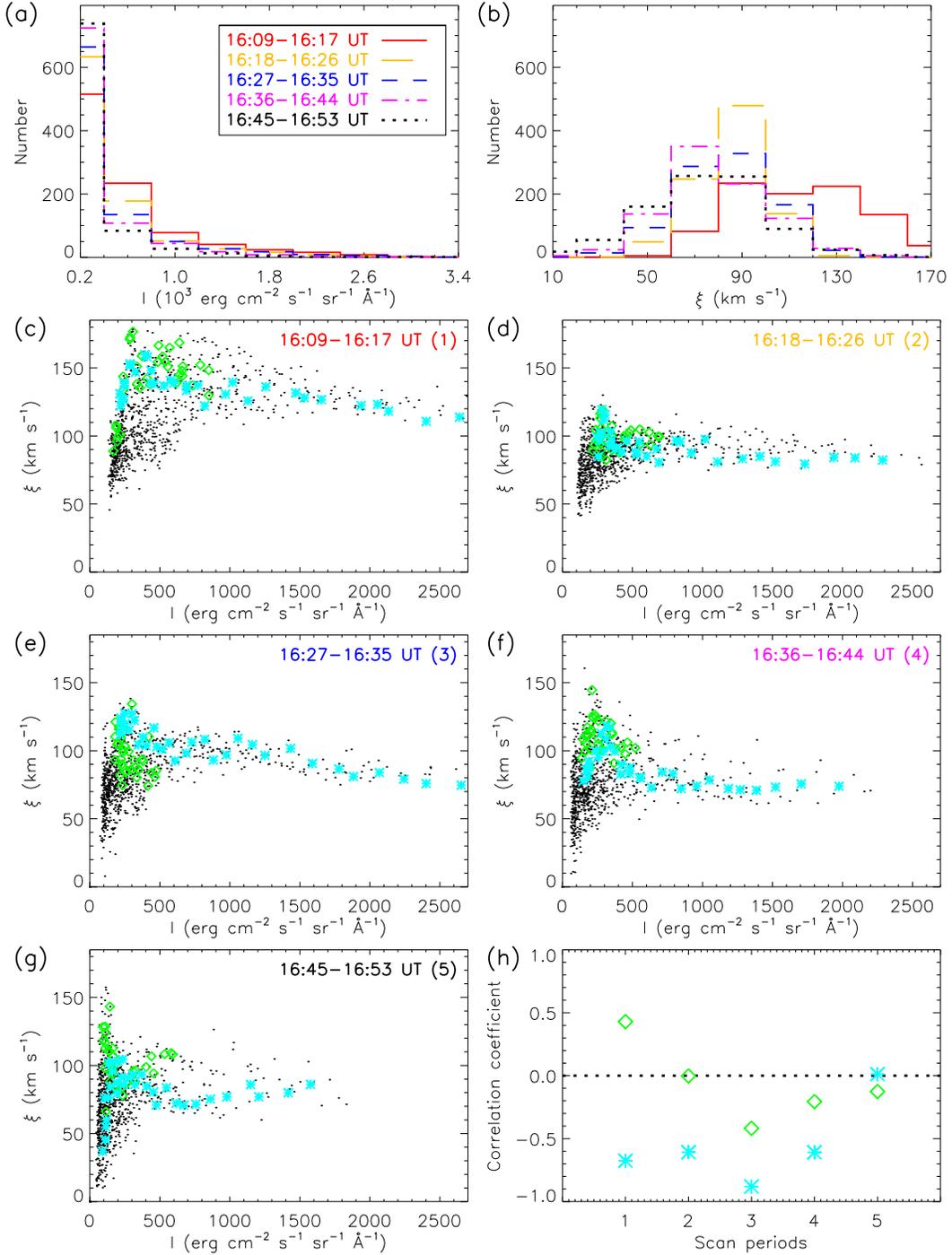}
\caption{{\small (a) and (b) Histograms of the \feiv~line intensity and nonthermal velocity within the magenta box marked in Figure \ref{fig-vel}. (c)--(g) Scatter plots of the \feiv~line intensity and nonthermal velocity within the magenta box (black dots) and at the locations marked by the green dotted line (green diamonds) and the cyan dashed-dotted line (cyan asterisks) in Figure \ref{fig-vel}. (h) Correlation coefficients between the line intensity and nonthermal velocity denoted by the green diamonds and cyan asterisks in panels (c)--(g).}}
\label{fig-htm}
\end{figure*}

Figure \ref{fig-vel} shows spatial and temporal variations of the line intensity and nonthermal velocity of \feiv~around the current sheet region. It is seen that the current sheet has very large nonthermal velocities ($>$150 km s$^{-1}$), especially during an early period of 16:09--16:17 UT when its emission was clearly visible in the EIS field-of-view. The nonthermal velocities in the current sheet are obviously larger than the ones ($\sim$100 km s$^{-1}$) at the cusp whose line intensities are relatively stronger. As the flare evolves, the nonthermal velocity in the current sheet decreases to 100 km s$^{-1}$ or below but is still larger than that in the surrounding region, and the line intensity there seems to decrease as well. In addition, we can see that the current sheet has a movement toward the south. Figures \ref{fig-htm}(a) and (b) give histograms of the line intensity and nonthermal velocity of \feiv~within the magenta box marked in Figure \ref{fig-vel} for different periods. One can clearly see the decay of nonthermal velocity from $>$100 km s$^{-1}$ to tens of km s$^{-1}$ and also a decrease in the line intensity.

From Figure \ref{fig-vel}, we also find that the largest nonthermal velocity is not cospatial with the strongest line intensity. It seems that most of the large nonthermal velocities are located at the edge of the current sheet rather than its central part. Here we use two straight lines, one tracking the axis of the current sheet with strong intensities (the cyan dashed-dotted line), and the other tracing some large nonthermal velocities at the edge of the current sheet (the green dotted line). The relationship between nonthermal velocity and line intensity at these locations is given in Figures \ref{fig-htm}(c)--(h). It is seen that a complex relationship exists. During the first four periods, the nonthermal velocity along the axis of the current sheet increases while the intensity decreases, with a correlation coefficient in the range between $-$0.61 and $-$0.88. In the last period, as well as for the case at the edge of the current sheet, there appears no good relationship between the nonthermal velocity and line intensity, with a correlation coefficient less than 0.5.

\subsection{Thickness of the Current Sheet Based on Line Intensity and Nonthermal Velocity Features}

The standard definition of the current sheet thickness is the FWHM of the current density distribution in the direction perpendicular to the axis of the current sheet, which is actually hard to be measured in observations \citep{linj15}. So an easier way or approximate method is using the brightness feature to determine the current sheet thickness. However, a thermal halo effect \citep{yoko97,seat09} or a thick plasma sheet \citep{liuy09,song12} surrounding the current sheet may affect the measurement of thickness. Here we use the line emission combined with nonthermal velocity features to derive the thickness of the current sheet.

\begin{figure*}
\centering
\includegraphics[width=14cm]{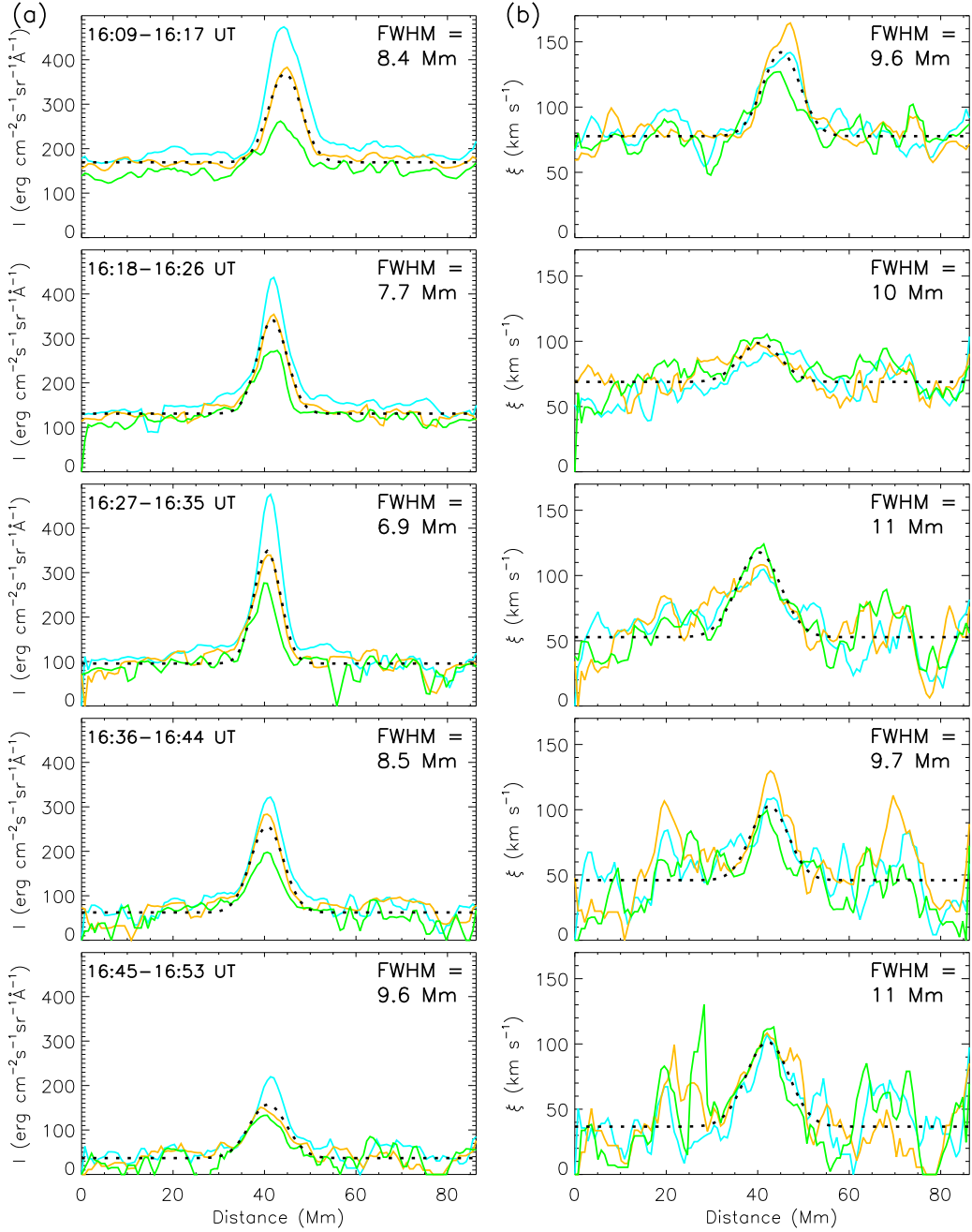}
\caption{{\small Temporal evolution of the \feiv~line intensity (a) and nonthermal velocity (b) distributions along the three vertical slices marked in Figure \ref{fig-vel}. The black dotted line in each panel is a Gaussian fitting to the average profile, whose FWHM is given.}}
\label{fig-wid}
\end{figure*}

Figure \ref{fig-wid} shows the \feiv~line intensity and nonthermal velocity distributions in the direction perpendicular to the axis of the current sheet (i.e., along the three vertical slices marked in Figure \ref{fig-vel}; at the heights of 115--132 Mm). Here we use a Gaussian function to fit the average profile and then adopt the FWHM as the current sheet thickness. It is seen that the thickness firstly decreases from 8.4 to 6.9 Mm and then increases to 9.6 Mm according to the emission feature. While the thickness derived from the nonthermal velocity feature is around 10 or 11 Mm at all times, which is slightly larger than the emission thickness. Note that the current sheet shows a clear isolation from the surrounding region in the emission profiles; while the nonthermal velocity profiles display some fluctuations or even small bumps around the current sheet particularly at late times, which might be related to plasma inflowing motions toward the current sheet or just caused by some noise because of low intensity in the \feiv~line.

\section{Summary and Discussions}
\label{sec-dis}

Using the high-resolution spectroscopic observations from \Hinode/EIS, we have presented the spatio-temporal variations of the line intensity and nonthermal velocity of \feiv~for the edge-on current sheet in an X8.2 flare. The long thin current sheet shows very large nonthermal velocities up to 200 km s$^{-1}$ at its early evolutionary stage. As the flare evolves, the nonthermal velocity in the current sheet decays to tens of km s$^{-1}$. The largest nonthermal velocity occurs at the edge of the current sheet, which is not cospatial with the greatest line intensity at the central part. Along the axis of the current sheet, the nonthermal velocity is negatively correlated with the line intensity in general. From the line intensity combined with nonthermal velocity distributions, we obtain a thickness in the range of 7--11 Mm for the current sheet. 

We note that \cite{warr17} also analyzed the same event using the EIS spectroscopic data. However, there are some differences between our study and theirs, both of which can supplement each other for the current sheet topic. (1) We focus on the relationship of nonthermal velocity and line intensity in the current sheet; and they studied the nonthermal velocity and in particular, the temperature of the current sheet. (2) We concentrate on the current sheet itself, i.e., extending to the higher corona with a heigh of $>$110 Mm; and they studied the current sheet mostly close to the cusp region. (3) They discussed their results in terms of plasma heating of the flare; and our results may provide some information on the nature of the current sheet and the physical processes that occur inside the current sheet. The latter issue is discussed in the following. 

Very large nonthermal velocities ($>$150 km s$^{-1}$) are detected in the current sheet, which are even greater than the values in the cusp region. This suggests that strong turbulence exists in the diffusion region \citep{ciar02,ciar08,bemp08,susi13,dosc14}. The turbulence is likely caused by plasma instabilities and/or by Petschek-type reconnection (e.g., \citealt{linj09,linj15}), with the magnetic energy cascading to small scales and various magnetic islands formed \citep{mcke13}. It should be mentioned that \cite{chen17} analyzed the same event and observed intermittent sunward outflow jets and anti-sunward blobs along the current sheet. All these are consistent with magnetic islands of different sizes being present, and small-scale magnetic reconnection possibly occurring inside the current sheet. This should help the magnetic energy dissipate more efficiently. 
 
We check the relationship between the turbulent motion and the hot emission of the current sheet. First, the turbulent velocity has a negative correlation with the line intensity at the central part of the current sheet, which seems to be also implied from the EIS observations in \cite{dosc14}. Second, the largest turbulent velocity is not cospatial with the greatest intensity but is found at the edge of the current sheet where the plasma is flowing into the current sheet. These may imply that the current sheet has internal fine structures, some of which contribute to large emission measure and some others to strong turbulence. The fine structures may exist in the form of magnetic islands that are in different radiative and dynamic phases. In fact, the difference in the spatial distributions of the line emission and turbulent motion in the current sheet seems not to be reported before. This is worth being investigated in more events as well as in numerical simulations in the future.

The thickness of current sheet, together with the correct approaches of measuring it, is still an important topic \citep{linj15}. Thanks to the high spatio-temporal resolution spectroscopic observations from EIS, we can obtain the current sheet thickness using the line emission as well as turbulent features. The derived thickness values (7--11 Mm) are in the range of the apparent values observed before and also way beyond the range expected by the traditional reconnection theory. However, our results imply that the cascading process reflected by strong turbulent motions may be crucial for accelerating the magnetic reconnection \citep{mcke13,linj15}, and that fast magnetic reconnection could proceed in such a thick current sheet in solar eruptions.

Overall, the excellent spectroscopic observations from \Hinode/EIS provide a good opportunity to study the nature of the current sheet and the physical processes therein. The results can be compared to numerical simulations of magnetic reconnection in solar eruption models.


\acknowledgments
\SDO~is a mission of NASA's Living With a Star Program. \Hinode~is a Japanese mission developed and launched by ISAS/JAXA. Y.L. thanks Dr. Jun Lin for valuable discussions. The authors are supported by NSFC under grants 11373023, 11403011, 11733003, 11233008, 11427803, and U1731241, by NKBRSF under grant 2014CB744203, and by the CAS Strategic Pioneer Program on Space Science under grants XDA15052200, XDA15320103, and XDA15320300. Y.L. is also supported by CAS Pioneer Hundred Talents Program, CAS Key Laboratory of Solar Activity of National Astronomical Observatories (KLSA201712), and by ISSI-BJ from the team ``Diagnosing Heating Mechanisms in Solar Flares".

\bibliographystyle{apj}

\begin{thebibliography}{}
\expandafter\ifx\csname natexlab\endcsname\relax\def\natexlab#1{#1}\fi

\bibitem[Bemporad(2008)]{bemp08}
 Bemporad, A.\ 2008, \apj, 689, 572 

\bibitem[Bemporad et al.(2006)]{bemp06}
 Bemporad, A., Poletto, G., Suess, S.~T., et al.\ 2006, \apj, 638, 1110

\bibitem[Cheng et al.(2017)]{chen17}
 Cheng, X., Li, Y., Wan, L.~F., et al.\ 2017, submitted

\bibitem[Ciaravella \& Raymond(2008)]{ciar08}
 Ciaravella, A., \& Raymond, J.~C.\ 2008, \apj, 686, 1372 

\bibitem[Ciaravella et al.(2002)]{ciar02}
 Ciaravella, A., Raymond, J.~C., Li, J., et al.\ 2002, \apj, 575, 1116

\bibitem[Ciaravella et al.(2013)]{ciar13}
 Ciaravella, A., Webb, D.~F., Giordano, S., \& Raymond, J.~C.\ 2013, \apj, 766, 65

\bibitem[Culhane et al.(2007)]{culh07}
 Culhane, J.~L., Harra, L.~K., James, A.~M., et al.\ 2007, \solphys, 243, 19

\bibitem[Doschek et al.(2014)]{dosc14}
 Doschek, G.~A., McKenzie, D.~E., \& Warren, H.~P.\ 2014, \apj, 788, 26
 
\bibitem[Ko et al.(2003)]{koyk03}
 Ko, Y.-K., Raymond, J.~C., Lin, J., et al.\ 2003, \apj, 594, 1068

\bibitem[Lemen et al.(2012)]{leme12}
 Lemen, J.~R., Title, A.~M., Akin, D.~J., et al.\ 2012, \solphys, 275, 17

\bibitem[Lin \& Forbes(2000)]{linj00}
 Lin, J., \& Forbes, T.~G.\ 2000, \jgr, 105, 2375

\bibitem[Lin et al.(2005)]{linj05}
 Lin, J., Ko, Y.-K., Sui, L., et al.\ 2005, \apj, 622, 1251
 
\bibitem[Lin et al.(2009)]{linj09}
 Lin, J., Li, J., Ko, Y.-K., \& Raymond, J.~C.\ 2009, \apj, 693, 1666

\bibitem[Lin et al.(2015)]{linj15}
 Lin, J., Murphy, N.~A., Shen, C., et al.\ 2015, \ssr, 194, 237
 
\bibitem[Linker et al.(2003)]{link03}
 Linker, J.~A., Miki{\'c}, Z., Lionello, R., et al.\ 2003, Physics of Plasmas, 10, 1971

\bibitem[Litvinenko(1996)]{litv96}
 Litvinenko, Y.~E.\ 1996, \apj, 462, 997

\bibitem[Liu et al.(2009)]{liuy09}
 Liu, Y., Luhmann, J.~G., Lin, R.~P., et al.\ 2009, \apjl, 698, L51
 
\bibitem[McKenzie(2013)]{mcke13}
 McKenzie, D.~E.\ 2013, \apj, 766, 39
 
\bibitem[Priest \& Forbes(2002)]{prie02}
 Priest, E.~R., \& Forbes, T.~G.\ 2002, \aapr, 10, 313
 
\bibitem[Reeves \& Golub(2011)]{reev11}
 Reeves, K.~K., \& Golub, L.\ 2011, \apjl, 727, L52
 
\bibitem[Savage et al.(2012)]{sava12}
 Savage, S.~L., McKenzie, D.~E., \& Reeves, K.~K.\ 2012, \apjl, 747, L40

\bibitem[Savage et al.(2010)]{sava10}
 Savage, S.~L., McKenzie, D.~E., Reeves, K.~K., Forbes, T.~G., \& Longcope, D.~W.\ 2010, \apj, 722, 329 
 
\bibitem[Schmieder et al.(2015)]{schm15}
 Schmieder, B., Aulanier, G., \& Vr{\v s}nak, B.\ 2015, \solphys, 290, 3457

\bibitem[Seaton et al.(2017)]{seat17} 
 Seaton, D.~B., Bartz, A.~E., \& Darnel, J.~M.\ 2017, \apj, 835, 139

\bibitem[Seaton \& Forbes(2009)]{seat09}
 Seaton, D.~B., \& Forbes, T.~G.\ 2009, \apj, 701, 348

\bibitem[Song et al.(2012)]{song12}
 Song, H.~Q., Kong, X.~L., Chen, Y., et al.\ 2012, \solphys, 276, 261

\bibitem[Susino et al.(2013)]{susi13}
 Susino, R., Bemporad, A., \& Krucker, S.\ 2013, \apj, 777, 93

\bibitem[Warren et al.(2017)]{warr17}
 Warren, H.~P., Brooks, D.~H., Ugarte-Urra, I., et al.\ 2017, arXiv:1711.10826
 
\bibitem[Warren et al.(2011)]{warr11}
 Warren, H.~P., O'Brien, C.~M., \& Sheeley, N.~R., Jr.\ 2011, \apj, 742, 92

\bibitem[Webb et al.(2003)]{webb03}
 Webb, D.~F., Burkepile, J., Forbes, T.~G., \& Riley, P.\ 2003, Journal of Geophysical Research (Space Physics), 108, 1440

\bibitem[Wood \& Neukirch(2005)]{wood05}
 Wood, P., \& Neukirch, T.\ 2005, \solphys, 226, 73

\bibitem[Yokoyama \& Shibata(1997)]{yoko97}
 Yokoyama, T., \& Shibata, K.\ 1997, \apjl, 474, L61

\end{thebibliography}

\end{document}